\documentclass[12pt]{iopart}

\usepackage{iopams}  
\usepackage{graphicx}
\begin{document}

\title[Phase stability of reversible polymers and nanospheres]{Phase stability of a reversible supramolecular polymer solution mixed with nanospheres}

\author{Remco Tuinier}

\address{Van $^{\prime}$t Hoff Laboratory, Debye Institute, Utrecht University, The Netherlands}
\address{DSM Research, ACES, P.O. Box 18, 6160 MD Geleen, The Netherlands}
\ead{remco.tuinier@dsm.com}
\begin{abstract}
Theory is presented for the phase stability of mixtures containing nanospheres and
non-adsorbing reversible supramolecular polymers. This was made possible by incorporating
the depletion thickness and osmotic pressure of reversible supramolecular polymer chains into
generalized free-volume theory, recently developed for investigating the phase behaviour of
colloidal spheres mixed with interacting polymers [\textit{Adv. Colloid Interface Sci.} \textbf{143} (2008) 1-47].
It follows that the fluid–fluid phase stability region where reversible
supramolecular polymer chains can be mixed with nanospheres is sensitive to the energy of
scission between the monomers and to the nanoparticle radius. One can then expect the
fluid–fluid coexistence curves to have a strong dependence on temperature and that shifting of
phase boundaries within a single experimental system should be possible by varying the
temperature. The calculations reveal the width of the stability region to be rather small. This
implies that phase homogeneity of product formulations containing reversible supramolecular
polymers is only possible at low nanoparticle concentrations.
\end{abstract}

\pacs{81.16.Fg, 64.75.Xc, 82.70.Dd}
\maketitle

\section{\label{intro}Introduction}
Reversible supramolecular polymers are polymer chains
that form due to spontaneous self-assembly of monomers \cite{Greef2008}.
Recent progress in supramolecular chemistry led to the development of several types of reversible supramolecular polymers
including hydrogen-bonded equilibrium polymers \cite{Esch2000,Brunsveld2001,Prins2001,Ikkala2002,Lehn2002} or coordination polymers that self-assemble via for instance metal ion-ligand group interactions \cite{Chen1994}.
Another class of such reversible polymers that behave similarly are worm-like surfactant micelles \cite{Cates1990}. Such supramolecular polymers can be used to form a large diversity of self-organized multifunctional materials \cite{Stupp1997}. 

\begin{figure}[ht]
\begin{center}
\includegraphics[width=9cm]{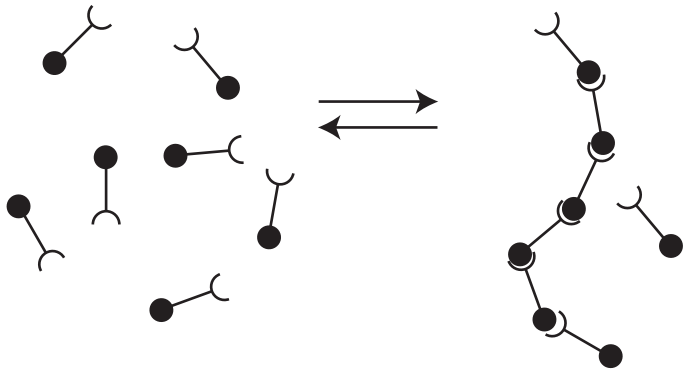}
\end{center}
\caption{Sketch of the formation of reversible supramolecular polymers composed of
individual monomers. By changing for instance temperature or monomer concentration
the equilibrium constant attains a different value leading to a modified chain length distribution.} \label{eqpols}
\end{figure}

The chain length (distribution) of these polymers is determined by the end-cap or scission energy and the polymer concentration \cite{Cates1990}. The scission energy, $U$, is the energy required to create two new chain ends. A schematic picture of the formation of these reversible polymers
is drawn in Fig.~\ref{eqpols}.
A larger monomer concentration and stronger attraction between the monomers enhance self-assembly into longer chains.
Often the bonds form and break on time scales that are accessible experimentally.
The fact that the bonds between the monomers are noncovalent
provides solutions of reversible supramolecular polymers with a chain length distribution that
depends on monomer concentration and temperature.  
Here a simple mean-field model \cite{Cates1990} is adopted to describe the equilibrium chain length distribution. For sake of simplicity only athermal chains are considered. For a given segment volume fraction $\varphi$, the segment volume fraction that has a chain length $M$,  $\varphi_M$, is then given by the exponential distribution:
\begin{equation}\label{N00}
\varphi_M \,=\,  \frac{\varphi}{{\left\langle M \right\rangle}} \,\ \textrm{e}^{-M/ \left\langle M \right\rangle} \quad \,\mbox{,}
\end{equation}
The brackets $\langle\cdot\rangle$ denote the ensemble average;
$\left\langle M \right\rangle$ is the number-averaged chain length given by:
\begin{equation}\label{Mav}
\left\langle M \right\rangle \,= \, \sqrt{\varphi} \,\ \textrm{e}^{U/2}\quad \,\mbox{,}
\end{equation}
where $U$ is the scission energy between the monomers (normalized by $kT$).
Following van der Gucht \textit{et al.} \cite{Gucht2004}, the
relative concentration of reversible supramolecular polymers
(further denoted as equilibrium polymers) $\phi_{\mathrm{p}}$ is defined as
\begin{equation}\label{phip}
\phi_{\mathrm{p}} = 
\varphi \left\langle M \right\rangle \quad \,\mbox{,}
\end{equation}
with 
$\phi_{\mathrm{p}} = 1$ defining the overlap concentration of equilibrium polymers
at which the equilibrium polymer coils exactly fill the total volume.

In this paper the effect of adding nanospheres to solutions of non-adsorbing equilibrium polymers is considered.
Such nanoparticles can add a certain functionality to the supramolecular polymer solution.
With the increasing interest in supramolecular polymers and
widening possibilities to synthesize them, the phase stability of such mixtures will become
increasingly relevant.
As far as I am aware, only a single (preliminary) experimental
study on the phase stability of nanoparticles mixed with (non-adsorbing) equilibrium polymers
has been reported \cite{Knoben2007}. Knoben \textit{et al}. examined 2,4-bis(2-ethylhexylureido)toluene
solutions mixed with stearylated silica nanospheres in cyclohexane and detected
a liquid-liquid phase separation at equilibrium polymer concentrations at and above 10 $g/L$ ($\varphi \gtrsim 0.01$).

\begin{figure}[ht]
\begin{center}
\includegraphics[width=10cm]{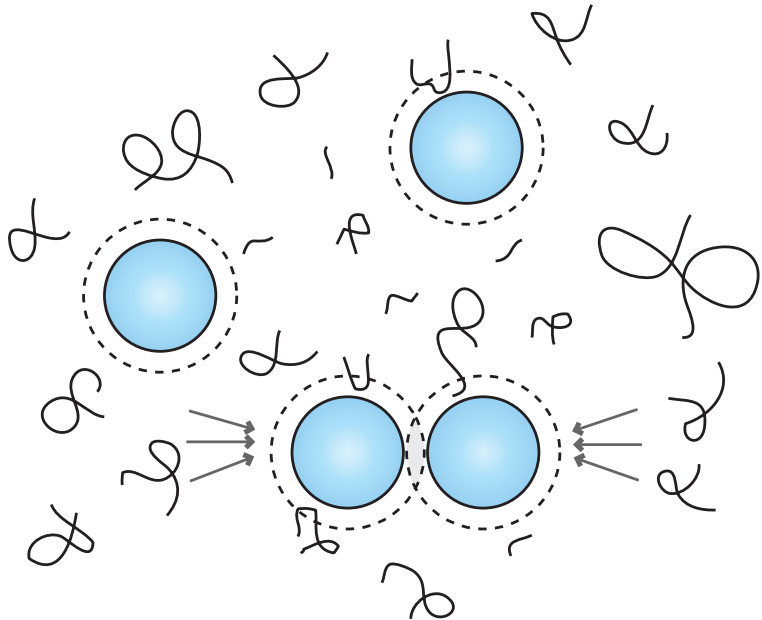}
\end{center}
\caption{Schematic picture of nanospheres in a solution containing non-adsorbing equilibrium polymers solution.
Note the broad chain length distribution of the equilibrium polymers.
The effective depletion layers are indicated by the
dashed circles. Upon overlap of depletion layers (the lower two spheres) the osmotic pressure acting upon the spheres is unbalanced.
The resulting attractive force is indicated by arrows.} \label{Depeqhs}
\end{figure}

The system of interest is sketched in Fig.~\ref{Depeqhs}. The equilibrium polymers are depleted from zones around the spheres termed depletion layers (dashed). In this paper, the mixture of nanospheres and equilibrium polymers in a solvent is studied using free volume theory.

\clearpage

In the early 1990s Lekkerkerker and others put forward free volume theory (FVT) for mixtures of spherical particles and (ideal) polymers \cite{Lekkerkerker1990,Lekkerkerker1992}. For solutions containing
non-adsorbing polymers, the nanoparticles are surrounded by depletion layers \cite{Asakura1954,Vrij1976,degennes1979}. Overlap of
depletion layers results in attraction between the particles \cite{depbook} and these attractive forces in turn induce phase transitions in nanoparticle plus polymer mixtures \cite{Ilett1995,Poon2002,Duijneveldt2007,Fleer2008,depbook}. 
The phase diagram topology depends on the polymer-to-sphere size ratio.

The power of FVT is that it is simple, insightful, \textit{and} accurate (see for instance \cite{Meijer1994,Bolhuis2002a,MonchoJorda2003,Dijkstra2006}). Over the years, FVT has been shown to be flexible in that it can be used to describe the phase behaviour of mixtures of colloidal rods and polymers \cite{Lekkerkerker1994}, dispersions of colloidal spheres and (small) colloidal rods \cite{Vliegenthart1999}, charged spheres mixed with polymer chains \cite{Fortini2005,Gogelein2008}, and mixtures of interacting polymers and colloidal spheres \cite{Aarts2002,Fleer2007b,Tuinier2008}.

Here I present generalized free volume theory for a mixture of spherical nanoparticles in a supramolecular polymer solution. The theory is outlined in section \ref{theory}, and the results for the phase stability are presented in section \ref{results}.

\section{\label{theory}Theory}

\noindent \subsection{\label{semi_grand_canonical_potential}
Semi-grand canonical potential}

Within generalized free-volume theory (GFVT), a nanoparticle-polymer mixture is
described by the semi-grand-canonical ensemble. The system of interest is held in osmotic equilibrium with a reservoir (r).
This reservoir is separated from the system via a hypothetical membrane that allows permeation of equilibrium polymers and solvent molecules  but which does not allow the nanospheres to enter the reservoir. The equilibrium polymer concentration in the system is set by the chemical
potential of the polymer chains in the reservoir, $\mu_{\mathrm{p}}^{\mathrm{r}}$. The system, with volume $V$ at temperature $T$, consists of $N_{\mathrm{n}}$ nanospheres with radius $a$ in a solvent with equilibrium polymers. The solvent can safely be considered as 'background' \cite{Lekkerkerker1990,LyklemaFICS4Ch5,Fleer2008}.
An expression is needed for the semi-grand potential $\Omega$ for the mixture of nanoparticles and equilibrium polymers in solution in order to be able to compute the thermodynamic properties of the mixture.
At given system volume and temperature, the normalized
semi-grand canonical potential, $\omega=\beta\Omega
v_{\mathrm{n}}/V$, can be written in normalized form as \cite{Aarts2002,Fleer2008}
\begin{equation}\label{reduced_semi_grand_canon_energy}
  \omega = f^{0}+ \omega_p \,\mbox{.}
\end{equation}
\noindent with
\begin{equation}
\label{MFVT} \omega_{\mathrm{p}} \;=\; -\;\beta v_{\mathrm{n}} \int_{0}^{\phi_{\mathrm{p}}^{\mathrm{r}}}
\alpha \left(\frac{\partial {\Pi}^{\mathrm{r}}}{\partial \phi_{\mathrm{p}}^{\mathrm{r}}} \right)
\textrm{d}\phi_{\mathrm{p}}^{\mathrm{r}} \quad ,
\end{equation}
where the reduced canonical free energy
$f^{0}=\beta F^{0} v_{\mathrm{n}}/V$. Here $1/\beta=k_{\mathrm{B}}T$ is
the thermal energy, $v_{\mathrm{n}}=4\pi a^3 /3$ is the volume
of a nanosphere, and $F^{0}$ is the Helmholtz energy of the pure nanoparticle dispersion.
The nanoparticle concentration is expressed in
the volume fraction $\eta = N_{\mathrm{n}}v_{\mathrm{n}}/V$.

In order to calculate the polymer concentration in the system one
needs the free-volume fraction $\alpha$ for the polymer chains in
the system, defined as
$\alpha(\eta,\mu_{\mathrm{p}}^{\mathrm{r}})=\langle
V_{\mathrm{free}}\rangle / V$, where $V_{\mathrm{free}}$ is the
free-volume in the system not occupied by the nanospheres and their
depletion zones. 

For the free volume fraction $\alpha$ the standard
scaled-particle result is used \cite{Reiss1959,Reiss1986,Reiss1992,Lekkerkerker1992,Aarts2002}:
\begin{equation}
\label{eqalfa} \alpha =(1-\eta) \exp(-A y - B y^2 - C
y^3),
\end{equation}
where $y = \eta/(1-\eta)$, $A=(1+{\delta / a})^{3}-1$, $B=3{(\delta / a)}^2
(\delta / a +3/2)$, and $C=3{(\delta / a)}^3$. Here $\delta$ is the thickness
of the depletion zone around a
nanosphere with radius $a$. From (\ref{MFVT}) and (\ref{eqalfa}) it follows the quantities
$\delta$ and $\partial \Pi /
\partial \phi_{\mathrm{p}}$ are needed to calculate $\omega_p$. These will be given below.

\noindent \subsection{\label{polymer contr} Reversible supramolecular polymer contribution}
The ensemble-averaged chain length $\left\langle M \right\rangle$ as a function of the normalized polymer concentration
$\phi_{\mathrm{p}}$ follows from inserting
(\ref{phip}) into (\ref{Mav}):
\begin{equation}\label{N0}
\left\langle M \right\rangle \, = \, \phi_{\mathrm{p}}^{1/3}
\quad\left({\textrm e}^{U/2}\right)^{2/3}\quad \,\mbox{.}
\end{equation}
The size of the equilibrium polymers $R_e$ is defined as:
\begin{equation}\label{R0}
R_{e} \,=\, \sqrt{\frac{\left\langle M \right\rangle}{6}} \quad \,\mbox{.}
\end{equation}
It is noted that all length scales are normalized
with the size of a polymer segment in this paper.

The depletion thickness for dilute and semi-dilute equilibrium polymer
solutions next to a flat wall, $\delta_{\mathrm{p}}$, was
obtained previously \cite{Gucht2004},
\begin{equation}\label{general_depletion_thickness_wall}
  \delta_{\mathrm{p}}\, =\, \frac{R_e}{\phi_{\mathrm{p}}}\left[ 2 \sqrt{1+2 \phi_{\mathrm{p}}} - \sqrt{2(\phi_{\mathrm{p}}+2)} \right]\,\mbox{.}
\end{equation}
In the dilute limit ($\phi_{\mathrm{p}}\to 0$),
$\delta_{\mathrm{p}}$ vanishes and in the
semi-dilute limit ($\phi_{\mathrm{p}} \gg 1$), the depletion thickness follows the mean-field scaling relation
$\delta_{\mathrm{p}}\sim \phi_{\mathrm{p}}^{-1/2}$ \cite{Gucht2004}.

\begin{figure}[ht]
\begin{center}
\includegraphics[width=9cm]{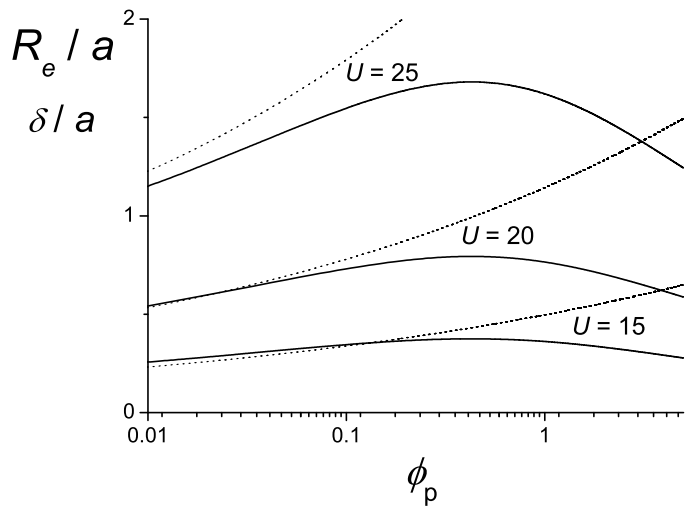}
\end{center}
\caption{Concentration dependence of the depletion thickness (solid curve) and chain size (dotted curve) of the equilibrium
polymers around a sphere for three values of the scission $U$ energy as indicated.} \label{figdelta}
\end{figure}

It is important to also account for curvature effects to accurately
describe the depletion thickness, $\delta$, around a spherical particle. In the mean-field case $\delta$ can be
approximated accurately by the simple power law \cite{Fleer2008}:
\begin{equation}\label{approx_curv_mean_field}
  \frac{\delta}{a} \, = \,0.842\, \left(\frac{\delta_{\mathrm{p}}}{a}\right)^{0.9}\,\mbox{.}
\end{equation}
In Fig. \ref{figdelta} the characteristic polymer size ($R_e$) and depletion thickness ($\delta$) of equilibrium polymers are plotted as a function of the relative polymer concentration $\phi_{\mathrm{p}}$ for a few values of the scission energy. It illustrates how the size of the equilibrium polymer chains ($R_e$) increases with $\phi_{\mathrm{p}}$. The depletion thickness also increases (as it follows $R_e$ in the dilute case) up to concentrations close to overlap (of order unity) above which $\delta$ decreases; in the semi-dilute polymer concentration regime $\delta$ follows the correlation length in a polymer solution \cite{degennes1979}. As the scission energy is larger, the chains will be larger at given relative polymer concentration $\phi_{\mathrm{p}}$. The same holds for the depletion thickness.

\begin{figure}[ht]
\begin{center}
\includegraphics[width=9cm]{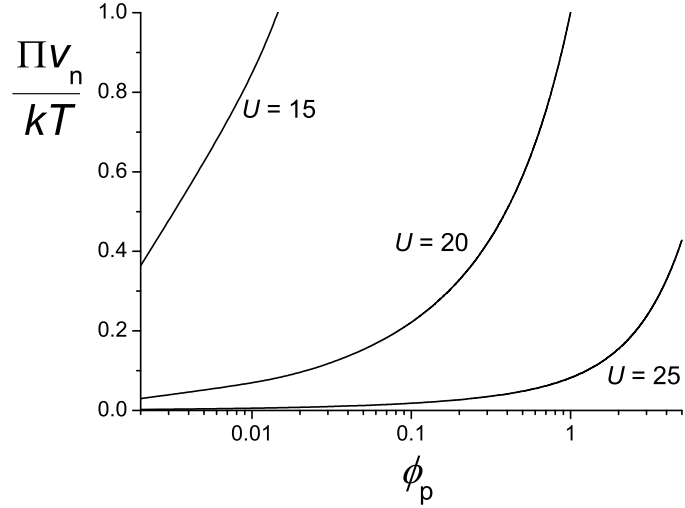}
\end{center}
\caption{Osmotic pressure of equilibrium
polymers for three values of the scission $U$ energy as indicated.} \label{figPi}
\end{figure}

For the osmotic pressure $\Pi$ of equilibrium polymers the following mean-field result is used:
\cite{Gucht2004}
\begin{equation}\label{general_osmotic_pressure}
\widetilde{P} = \beta v_{\mathrm{p}} \Pi = \phi_{\mathrm{p}} + \frac{1}{2} \phi_{\mathrm{p}}^{2} \quad \,\mbox{.}
\end{equation}
Here $\widetilde{P}$ is the normalized osmotic pressure.
The polymer concentration dependence of osmotic pressure is plotted in Fig. \ref{figPi} for various values of the scission energy $U$.
The derivative of osmotic pressure with respect to polymer concentration is
\begin{equation}\label{dP_mf}
 \frac{\partial \widetilde{P}}{\partial \phi_{\mathrm{p}}}  = \beta v_{\mathrm{n}} \frac{\partial \Pi}{\partial \phi_{\mathrm{p}}} \,=\, \left(\frac{a}{R_e}\right)^{3}\left[1 + \phi_{\mathrm{p}} \right]\quad \,\mbox{.}
\end{equation}

\noindent \subsection{\label{colloid contr} Free energy of a
nanoparticle dispersion of hard spheres}

The canonical free energy of a fluid dispersion containing only hard spheres can
be accurately described using:
\begin{equation}
f^{0}
 = \eta \ln( \eta
\Lambda^3 / v_{\mathrm{n}}) - \eta \;+\;
\frac{4{\eta}^2 - 3{\eta}^3}{(1-{\eta})^2} \quad \mbox{,}
\end{equation}
\noindent where $\Lambda=h/\sqrt{2\pi m_{\mathrm{n}}
k_{\mathrm{B}}T}$ is the thermal wavelength, $m_{\mathrm{n}}$
is the nanoparticle mass and $h$ is Planck's constant. The first two terms on the
right-hand side of this equation comprise the ideal contribution,
whereas the last term describes the hard-sphere interaction, which follows the equation of state given by Carnahan and Starling \cite{Carnahan1969}.

Standard thermodynamics enables computation of the osmotic pressure of the nanoparticle dispersion and the chemical potential of the nanoparticles.
The chemical potential $\widetilde{\mu}^{0}$ = $\mu^{0}/kT$ = $\partial
f^{0} / \partial \eta$ follows as:
\begin{equation} \label{eqmuCS}
\widetilde {\mu}^{0} = \ln \frac
{\Lambda^3}{v_{\mathrm{n}}} + \ln \eta + \displaystyle{\frac{(8-9 \eta + 3
\eta^2)\eta}{(1-\eta)^3}} \quad .
\end{equation}
The dimensionless pressure $\widetilde{P}^{0}$ = $P_{\mathrm{n}}v_{\mathrm{n}}/kT$ can subsequently
be computed from $\widetilde{P}^{0} =
\eta \widetilde{\mu}^{0} - f^{0}$, giving
\begin{equation} \label{eqPCS}
\widetilde{P}^{0} = \frac{\eta + \eta^2 +
\eta^3-\eta^4 }{(1- \eta)^3} \quad ,
\end{equation}
the well-known Carnahan-Starling pressure.
 
\noindent \subsection{\label{binodals} Fluid-fluid binodals}
All ingredients are now available to calculate the fluid-fluid phase behaviour of the nanoparticle-equilibrium polymer
mixture by solving the coexistence equations
\begin{equation} \label{eqbinmu}
\widetilde{\mu}_{\mathrm{n}}^I = \widetilde{\mu}_{\mathrm{n}}^{II} \quad,
\end{equation}
and
\begin{equation} \label{eqbinP}
\widetilde{P}_{\mathrm{tot}}^{I} = \widetilde{P}_{\mathrm{tot}}^{II}  
\end{equation}
for the chemical potential of the nanospheres and the total osmotic pressure of a phase $I$ in equilibrium with a phase $II$.

The phase diagrams for hard spheres
plus interacting polymers using the general expression (\ref{reduced_semi_grand_canon_energy}) and its ingredients
now follow from computing the chemical potential $\widetilde{\mu}_{\mathrm{n}} = (\partial
\omega / \partial \eta)$ and total pressure $\widetilde{P}_{\mathrm{tot}} =
\eta \widetilde{\mu}_{\mathrm{n}} - \omega$,
\begin{equation} \label{eqRT4H}
\widetilde{\mu}_{\mathrm{n}}   = \widetilde{\mu}^0  +
\int_0^{\phi_{\mathrm{p}}^r} g  \left( \frac{\partial
\widetilde{P}^r}{\partial \phi_{\mathrm{p}}^{R'}} \right)
\mathrm{d} \phi_{\mathrm{p}}^{R'} \quad ,
\end{equation}
\begin{equation} \label{eqRT4J}
\widetilde{P}_{\mathrm{tot}}  =  \widetilde{P}^{0}   +
\int_0^{\phi_{\mathrm{p}}^r} h  \left( \frac{\partial
\widetilde{P}^r}{\partial \phi_{\mathrm{p}}^{R'}} \right)
\mathrm{d} \phi_{\mathrm{p}}^{R'} \quad ,
\end{equation}
with $g$ and $h$ given by:
\begin{equation} \label{eq3p41}
g = e^{-Q} \left\{ 1 + [1+ y](A + 2B y + 3Cy^2)
\right\}
\end{equation}
and
\begin{equation} \label{eq3p42}
h = e^{-Q} \left\{  1 + A y + 2B y^2 + 3C y^3 \right\}
\quad.
\end{equation}
with
$A$, $B$ and $C$ defined in (and below) equation (\ref{eqalfa}) and where $Q$ is given by
\begin{equation} \label{Qdef}
Q = A y + B y^2 + C y^3.
\end{equation}
The normalized osmotic pressure of the equilibrium polymers in the reservoir is
given by Eq. (\ref{general_osmotic_pressure}) and the osmotic compressibility by Eq. (\ref{dP_mf}).
Coexistence curves then follow from Eq. (\ref{eqbinmu}) and Eq. (\ref{eqbinP}) by using the common tangent construction.

\begin{figure}[ht]
\begin{center}
\includegraphics[width=11cm]{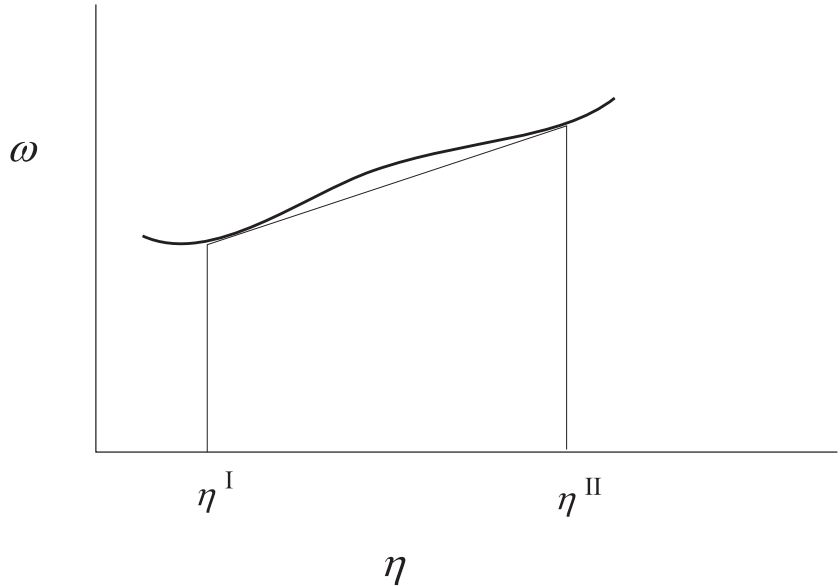}
\end{center}
\caption{Dimensionless semi-grand potential $\omega$ as a function of nanoparticle volume fraction $\eta$. The common tangent construction
identifies the binodal compositions $\eta^{I}$ and $\eta^{II}$. In this way one can determine the phase coexistence in mixtures of hard nanospheres and equilibrium polymers.} \label{figomega}
\end{figure}
In Fig.~\ref{figomega} it is demonstrated how the (normalized) semi-grand potential depends on nanosphere particle volume
fraction under conditions where the fluid phase exhibits an instability as follows from the local maximum in $\omega$.
Such a curve depends on the reservoir equilibrium polymer concentration, the nanoparticle size and on the scission energy. The common tangent construction in Fig.~\ref{figomega} allows
to determine the compositions of the two coexisting phases. From these binodals the full phase diagram can be constructed.


\section{\label{results}Results}

\begin{figure}[ht]
\begin{center}
\includegraphics[width=14cm]{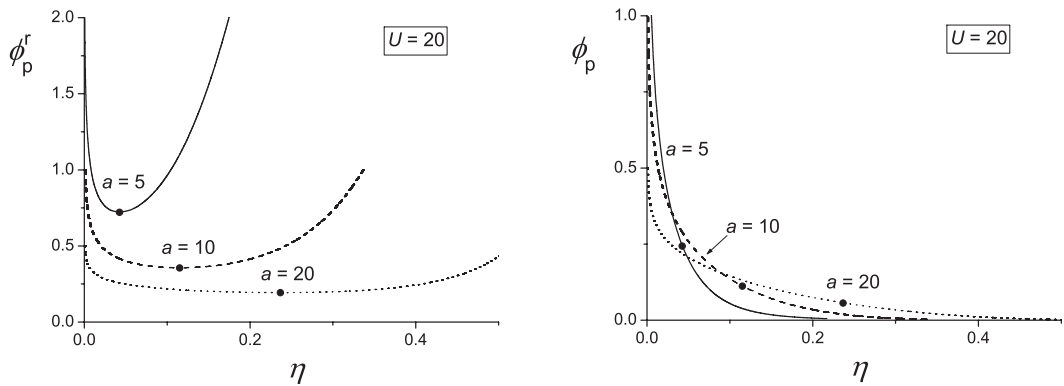}
\end{center}
\caption{Fluid-fluid coexistence curves of equilibrium polymers mixed with nanospheres at fixed scission energies $U$ = 20 for various values of the nanoparticle radius $a$ =5 (solid curve),
$a$ = 10 (dashed curve) and $a$ = 20 (dotted curve). Dots are the critical points.} \label{figphaseradius}
\end{figure}

In Fig.~\ref{figphaseradius} fluid-fluid binodal curves are presented for nanospheres of $a$ = 5, 10 and 20 with sizes given in units of the equilibrium polymer monomer size. The scission energy here was fixed at 20 ($kT$).
The left panel represents the relative polymer concentration in the reservoir ($\phi_p^r$) on the ordinate versus the nanoparticle volume fraction ($\eta$) on the abscissa. In this representation,
it follows that relatively larger polymer chains (smaller $a$) shift the fluid-fluid binodal curves to higher relative polymer concentrations. It is clear that the binodal sensitively depends on $a$. The relative polymer concentration in the system, $\phi_p$, follows from $\phi_p = \alpha \phi_{p}^{r}$. The resulting binodal curves are quite close to each other (right panel in Fig.~\ref{figphaseradius}). This is due to the fact that the free volume fraction depends on the relative polymer concentration and passes through a maximum around $\phi_{\mathrm{p}}$ = 0.6. This result shows that the stability region is rather limited for mixtures of nanospheres and equilibrium polymers.

\begin{figure}[ht]
\begin{center}
\includegraphics[width=14cm]{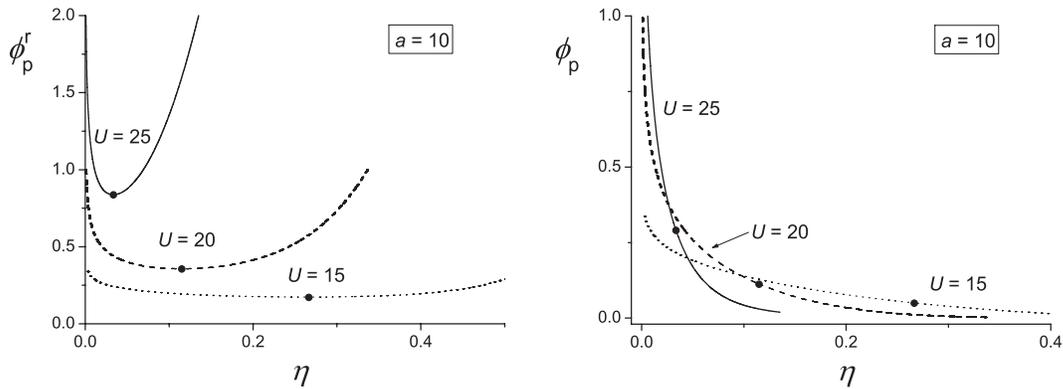}
\end{center}
\caption{Fluid-fluid coexistence curves of equilibrium polymers mixed with nanospheres of radius $a$ = 10 for scission energies $U$ = 15 (dotted curve),
$U$ = 20 (dashed curve) and $U$ = 25 (solid curve). Dots: critical points.} \label{figphaseU}
\end{figure}
In Fig.~\ref{figphaseU} fluid-fluid binodal curves are shown for mixtures of nanospheres with fixed size $a$ = 10 mixed with equilibrium polymers for various values of the scission energy $U$, the change of which is experimentally accessible via manipulation of temperature. It follows that increasing $U$ leads to a
significant shift upwards of the fluid-fluid binodals when the relative equilibrium polymer concentration is plotted in the reservoir representation
(left panel of Fig.~\ref{figphaseU}).
A larger value for $U$ leads to a larger average equilibrium polymer size. Therefore these results are in agreement with (G)FVT results for (normal) polymer-sphere mixtures \cite{Lekkerkerker1992,Ilett1995,Aarts2002}. Conversion towards system polymer concentrations
as in practice (right panel of Fig.~\ref{figphaseU}) shows the stability region is small and the binodal curves are close.

\begin{figure}[ht]
\begin{center}
\includegraphics[width=9cm]{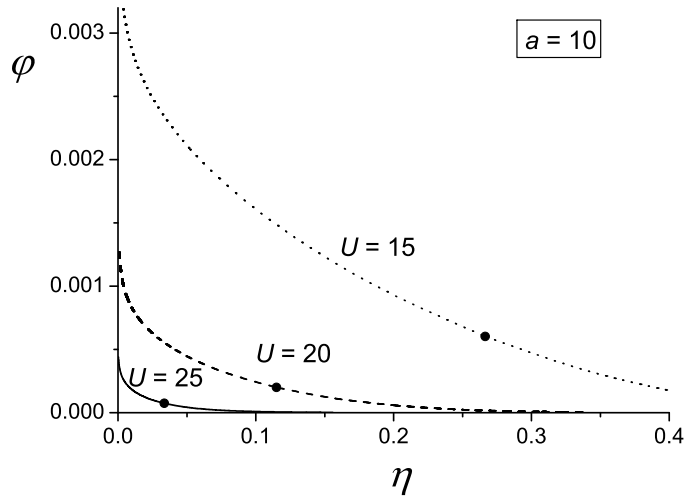}
\end{center}
\caption{Fluid-fluid coexistence curves of equilibrium polymers in system segment concentrations $\varphi$ mixed with nanospheres of radius $a$ for scission energies $U$ = 15 (dotted curve),
$U$ = 20 (dashed curve) and $U$ = 25 (solid curve). Dots are the critical points.} \label{figphaseUsegm}
\end{figure}

What is observed in practice? The relative polymer concentration is insightful since it identifies the concentration regime; dilute regime for $\phi_{\mathrm{p}}$ $<$ 1, semi-dilute regime if $\phi_{\mathrm{p}}$ $>$ 1. The quantity $\phi_{\mathrm{p}}$ is very sensitive to the scission energy $U$, while the segment concentration $\varphi$ is of course unaffected. The phase diagram of Fig.~\ref{figphaseU} (right panel) is now plotted in terms of the segment concentration on the ordinate, as represented in Fig.~\ref{figphaseUsegm}. It follows that variations of $U$ can have a dramatic effect on the binodal fluid-fluid coexistence curves observed in experiment.
Changing the relative size of polymer and spherical particle for mixtures of (normal) polymers mixed with nanoparticles requires significant experimental effort; one needs a different system with either particles or polymers of another size. This differs from changing $U$ in mixtures of reversible supramolecular polymers and nanospheres in the sense that one only needs to modify the temperature. With a single mixture one can then scan through the phase diagram by diluting, concentrating and changing the temperature.
No quantitative comparison can yet be performed with experimental work. The single experimental study that is available \cite{Knoben2007} used chain stoppers, which means we can not directly compare with their system. Further, their observations of (in)stability were quite limited. It will be interesting to compare the theory presented in this work to future phase diagrams on mixtures of nanoparticles and equilibrium polymer chains.

\begin{figure}[ht]
\begin{center}
\includegraphics[width=9cm]{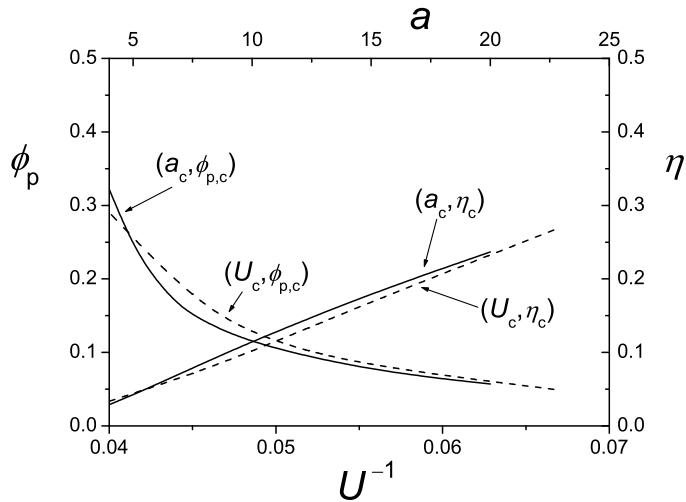}
\end{center}
\caption{Critical curves for $\phi_{\mathrm{p}}$ and $\eta$ as a function of $a$ and $1/U$ demonstrating their effects lead to similar shifts of the fluid-fluid binodal curves.} \label{figcrit}
\end{figure}

No remark was made yet on the critical point. For short-range attractions it is known that mean-field theories incorrectly predict the critical region,
whereas long-ranged attractions are well described. In Figures \ref{figphaseradius} and \ref{figphaseU}
it can be observed that increasing $U$ and decreasing $a$ imposes similar effects on the critical points and the binodals.
These critical $\phi_{\mathrm{p}}$ and $\eta$ values are plotted as a function of $a$ and $U^{-1}$ in a single plot in Fig.~\ref{figcrit}. It follows that, when scaled properly, the influence of $a$ and $U^{-1}$ are very similar and that one can modify the binodals in a similar fashion via adjusting $a$ or $U$.

\section{\label{concl}Concluding remarks}
Theory was presented for the fluid-fluid phase transition of mixtures of dispersed nanospheres
in solution containing equilibrium polymers. This was based upon generalized free volume theory originally developed for spherical particles mixed with interacting monodisperse polymer chains. The equilibrium polymers in solution were treated as athermal chains using a mean-field model. Within this model analytical expressions are available for the size, chain length (distribution), osmotic pressure and depletion thickness. In the dilute monomer concentration regime, the depletion thickness increases with monomer concentration while depletion thickness decreases in the semi-dilute equilibrium polymer concentration regime. A key parameter that sets the averaged chain length of the polymers is the scission energy; the energy required to create two new chain ends. In practice the scission energy can be varied by adjusting the temperature. An increase in scission energy results in increases in depletion thickness, while the osmotic pressure decreases at given relative polymer concentration.

The generated fluid-fluid binodal curves can partly be interpreted using previous findings from (generalized) free volume theory. These results can provide new insight into experimental observations. Upon varying the scission energy via the temperature one can, in combination with dilution and concentration of the dispersion, scan through the scission energy-dependent phase diagram. Finally, it was found that the critical point scales in a similar way with scission energy as with the inverse of the nanosphere radius.

In a later contribution, the effect of polymer solvency, as well as an account of the fluid-solid equilibrium, is planned.

\ack {John Zupancich, DSM Research and Jasper van der Gucht, Wageningen University are acknowledged for reading and commenting the manuscript. Further, I wish to express my appreciation to Mieke Kr{\"o}ner for her support and preparation of some of the figures.}
\\



\section*{References}
\begin{thebibliography}{10}

\bibitem{Greef2008}
T.F.A. de~Greef and E.W. Meijer.
\newblock {\em Nature}, 453:171, 2008.

\bibitem{Esch2000}
J.H. Van~Esch and B.L. Feringa.
\newblock {\em Angewandte Chemie}, 39:2263, 2000.

\bibitem{Brunsveld2001}
L.~Brunsveld, B.J.B. Folmer, E.W. Meijer, and R.P. Sijbesma.
\newblock {\em Chem. Rev.}, 101:4071, 2001.

\bibitem{Prins2001}
L.J. Prins, D.N. Reinhoudt, and P.~Timmerman.
\newblock {\em Angew. Chem.}, 40:2382, 2001.

\bibitem{Ikkala2002}
O.~Ikkala and G.~Ten~Brinke.
\newblock {\em Science}, 295:2407, 2002.

\bibitem{Lehn2002}
J.-M. Lehn.
\newblock {\em PNAS}, 99:4763, 2002.

\bibitem{Chen1994}
H.~Chen, J.A. Cronin, and R.D. Archer.
\newblock {\em Macromolecules}, 27:2174, 1994.

\bibitem{Cates1990}
M.E. Cates and S.J. Candau.
\newblock {\em J. Phys.: Condens. Matter}, 2:6869, 1990.

\bibitem{Stupp1997}
S.I. Stupp, V.~LeBonheur, K.~Walker, L.S. Li, K.E. Huggins, M.~Keser, and
  A.~Amstutz.
\newblock {\em Science}, 276:384, 1997.

\bibitem{Gucht2004}
J.~van~der Gucht, N.A.M. Besseling, and G.J. Fleer.
\newblock {\em Macromolecules}, 37:3026, 2004.

\bibitem{Knoben2007}
W.~Knoben, N.A.M. Besseling, and M.A. Cohen~Stuart.
\newblock {\em Langmuir}, 23:6095, 2007.

\bibitem{Lekkerkerker1990}
H.N.W. Lekkerkerker.
\newblock {\em Colloids and Surfaces}, 51:419, 1990.

\bibitem{Lekkerkerker1992}
H.N.W. Lekkerkerker, W.C.K. Poon, P.N. Pusey, A.~Stroobants, and P.~B. Warren.
\newblock {\em Europhys. Lett.}, 20:559, 1992.

\bibitem{Asakura1954}
S.~Asakura and F.~Oosawa.
\newblock {\em J. Chem. Phys.}, 22:1255, 1954.

\bibitem{Vrij1976}
A.~Vrij.
\newblock {\em Pure Appl. Chem.}, 48:471, 1976.

\bibitem{degennes1979}
P.G. De~Gennes.
\newblock {\em Scaling Concepts in Polymer Physics}.
\newblock Cornell University Press, Ithaca, 1979.

\bibitem{depbook}
H.N.W.~Lekkerkerker and R.~Tuinier.
\newblock {\em Colloids and the Depletion Interaction}.
\newblock Springer, Dordrecht, 2011.

\bibitem{Ilett1995}
S.~M. Ilett, A.~Orrock, W.~C.~K. Poon, and P.~N. Pusey.
\newblock {\em Phys. Rev. E}, 51:1344, 1995.

\bibitem{Poon2002}
W.C.K. Poon.
\newblock {\em J. Phys: Condens. Matter}, 14:R859, 2002.

\bibitem{Duijneveldt2007}
J.S. Van~Duijneveldt, K.~Mutch, and J.~Eastoe.
\newblock {\em Soft Matter}, 3:155, 2007.

\bibitem{Fleer2008}
G.J. Fleer and R.~Tuinier.
\newblock {\em Adv. Colloid Interface Sci.}, 143:1--47, 2008.

\bibitem{Meijer1994}
E.~J. Meijer and D.~Frenkel.
\newblock {\em J. Chem. Phys.}, 100:6873, 1994.

\bibitem{Bolhuis2002a}
P.~G. Bolhuis, A.~A. Louis, and J.~P. Hansen.
\newblock {\em Phys. Rev. Lett.}, 89:128302, 2002.

\bibitem{MonchoJorda2003}
A.~Moncho-Jorda, A.A. Louis, P.G. Bolhuis, and R.~Roth.
\newblock {\em J. Phys.: Condens. Matter}, 15:S3429, 2003.

\bibitem{Dijkstra2006}
M.~Dijkstra, R.~van Roij, R.~Roth, and A.~Fortini.
\newblock {\em Phys. Rev. E}, 73:041409, 2006.

\bibitem{Lekkerkerker1994}
H.N.W. Lekkerkerker and A.~Stroobants.
\newblock {\em Il Nuovo Cimento}, D16:949, 1994.

\bibitem{Vliegenthart1999}
G.A. Vliegenthart and H.~N.~W. Lekkerkerker.
\newblock {\em J. Chem. Phys.}, 111:4153, 1999.

\bibitem{Fortini2005}
A.~Fortini, M.~Dijkstra, and R.~Tuinier.
\newblock {\em J. Phys.: Condens. Matt.}, 17:7783--7803, 2005.

\bibitem{Gogelein2008}
C.~G{\"o}gelein and R.~Tuinier.
\newblock {\em Eur. Phys. J. E.}, 27:171, 2008.

\bibitem{Aarts2002}
D.G.A.L. Aarts, R.~Tuinier, and H.N.W. Lekkerkerker.
\newblock {\em J. Phys: Condens. Matter}, 14:7551, 2002.

\bibitem{Fleer2007b}
G.J. Fleer and R.~Tuinier.
\newblock {\em Phys. Rev. E}, 76:041802, 2007.

\bibitem{Tuinier2008}
R.~Tuinier, P.A. Smith, W.C.K. Poon, S.U. Egelhaaf, D.G.A.L. Aarts, H.N.W.
  Lekkerkerker, and G.J. Fleer.
\newblock {\em Europhys. Lett.}, 82:68002, 2008.

\bibitem{LyklemaFICS4Ch5}
A.~Vrij and R.~Tuinier.
\newblock {\em Chapter 5 in: Fundamentals in Colloid and Interface Science,
  Vol. 4, J. Lyklema (Ed.).}
\newblock Elsevier, Amsterdam, 2005.

\bibitem{Reiss1959}
H.~Reiss, H.L. Frisch, and J.L. Lebowitz.
\newblock {\em J. Chem. Phys.}, 31:369, 1959.

\bibitem{Reiss1986}
H.~Reiss and A.D. Hammerich.
\newblock {\em J. Phys. Chem.}, 90:6252, 1986.

\bibitem{Reiss1992}
H.~Reiss.
\newblock {\em J. Phys. Chem.}, 96:4736, 1992.

\bibitem{Carnahan1969}
N.F. Carnahan and K.E. Starling.
\newblock {\em J. Chem. Phys.}, 51:635, 1969.

\end{thebibliography}
\end{document}